\documentclass[
superscriptaddress,
 amsmath,amssymb,
 aps,
 prl,
 floats,
 twocolumn,
 epsf,
]{revtex4-2}

\usepackage{mathtools}% Better and more beautiful mathematical formulas (loads amsmath)
\usepackage{xcolor}% All the colors
\usepackage{graphicx}% Include figure files
\usepackage{dcolumn}% Align table columns on decimal point
\usepackage{bm}% bold math
\usepackage{bbm}
\usepackage{physics}%
\usepackage{booktabs}% More beautiful tables with \toprule, \midrule, \bottomrule, ...

\usepackage[colorlinks=true,linkcolor=blue,citecolor=red]{hyperref}
\usepackage{tikz}
\usepackage[compat=1.1.0]{tikz-feynhand} % for feynman graphs
\usepackage{textgreek}
\usepackage[capitalise]{cleveref}% fancy references
\usepackage[normalem]{ulem} % to strike out text

\newcommand{\ii}{\mathrm{i}}
\newcommand{\e}{\mathrm{e}}
\newcommand{\pdg}{{\ensuremath{\phantom{\dagger}}}}

\newcommand{\qv}{\ensuremath{\mathbf{q}}}

\newcommand{\pp}{\ensuremath{{pp}}}
\newcommand{\ph}{\ensuremath{{ph}}}

\renewcommand{\Tr}{\operatorname{Tr}}
\renewcommand{\Re}{\operatorname{Re}}

\DeclareRobustCommand\fermion{
    \begin{tikzpicture}[]
        \begin{feynhand}
         \vertex (a) at (-0.5,0);
         \vertex (b) at (0.5,0);
        \propag [fermion] (a) to (b);
        \end{feynhand}
        \end{tikzpicture}
}
\DeclareRobustCommand\anfermion{
    \begin{tikzpicture}[]
        \begin{feynhand}
        \vertex (a) at (-0.5, 0.);
        \vertex (b) at (0.5,0.);
        \propag [antfer] (a) to (b);
        \end{feynhand}
        \end{tikzpicture}
}

\begin{document}

\author{M.~Reitner}
\affiliation{Institute of Solid State Physics, TU Wien, 1040 Vienna, Austria}
\author{L.~Crippa}
\affiliation{Institut f\"ur Theoretische Physik und Astrophysik and W\"urzburg-Dresden Cluster of Excellence ct.qmat, Universit\"at W\"urzburg, 97074 W\"urzburg, Germany}
\author{D.~R.~Fus}
\affiliation{Institute of Solid State Physics, TU Wien, 1040 Vienna, Austria}
\author{J.~C.~Budich}
\affiliation{Institute of Theoretical Phyiscs, Technische Universität Dresden and W\"urzburg-Dresden Cluster of Excellence ct.qmat, 01062 Dresden, Germany}
\affiliation{Max Planck Institute for the Physics of Complex Systems, N\"othnitzer Str. 38, 01187 Dresden, Germany}
\author{A.~Toschi}
\affiliation{Institute of Solid State Physics, TU Wien, 1040 Vienna, Austria}
\author{G.~Sangiovanni}
\affiliation{Institut f\"ur Theoretische Physik und Astrophysik and W\"urzburg-Dresden Cluster of Excellence ct.qmat, Universit\"at W\"urzburg, 97074 W\"urzburg, Germany}

\title{Protection of Correlation-Induced Phase Instabilities by Exceptional Susceptibilities}

\date{\today}

\begin{abstract}
At thermal equilibrium, we find that generalized susceptibilities encoding the static physical response properties of Hermitian many-electron systems possess inherent non-Hermitian (NH) matrix symmetries. This leads to the generic occurrence of exceptional points (EPs), i.e., NH spectral degeneracies, in the generalized susceptibilities of prototypical Fermi-Hubbard models, as a function of a single parameter such as chemical potential. We demonstrate that these EPs are necessary to promote correlation-induced thermodynamic instabilities, such as phase-separation occurring in the proximity of a Mott transition, to a topologically stable phenomenon.
\end{abstract}

\maketitle

\vskip 5mm

\noindent
{\sl Introduction.} Topology has been introduced in physics to understand robustness. A pioneering achievement of this approach is the understanding of the quantum Hall effect~\cite{vonklitzing1980}, where the striking quantization of a transverse conductance has been explained in terms of topological properties of Bloch bands~\cite{laughlin1981,halperin1982,TKNN1982}. Since then, topology has conquered a wide range of physical settings far beyond the band theory of solids~\cite{bernevig2013,hasan2010,qizhang2011,wen2010,huber2016,bradlyn2017,ozawa2019,gong2018,song2020,xue2022,bergholtz2021}. As a novel direction within this paradigm, here we explain the robustness of phase instabilities in Fermi-Hubbard models by revealing and studying unique topological properties of their {\emph{generalized susceptibilities}}~\cite{rohringer2012local,thunstrom2018analytical}. These quantities, which encode information on the electronic fluctuations of a many-body system, can be linked to basic observables such as the uniform charge response (isothermal compressibility) $\chi_{\boldsymbol{q}=0}$~\cite{bickers2004self,rohringer2012local}.

\begin{figure}[htp]
\includegraphics[width=0.4\textwidth]{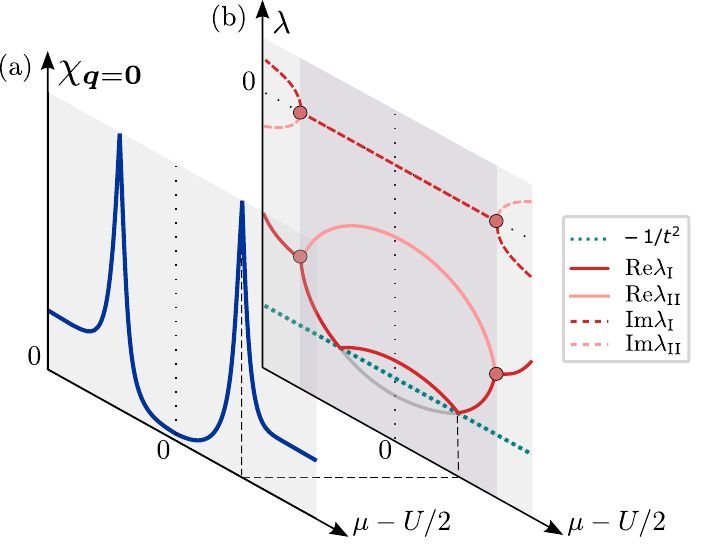} 
\caption{\label{fig:schematic}
Schematic illustration of the enhancement of the uniform charge response $\chi_{\bm{q}=\bm{0}}$  (a) triggered by the eigenvalue $\lambda_{\text{I}}$ reaching the condition $\lambda_{\text{I}}=-1/t^2$ (b). The chemical potential $\mu$ at which the divergence happens is located within a finite range, bounded by the EPs (red dots on (b)). If a divergence is found, no small perturbation can wash it out, as the intersection between the ``rectified'' $\lambda_{I}$ curve (thin gray line) and the $-1/t^{2}$ limit can only be continuously shifted. }
\end{figure}

Importantly, even for closed many-body systems described by a Hermitian Hamiltonian, such generalized susceptibilities naturally acquire a complex spectrum as matrices in Matsubara frequency space, and are thus subject to a non-Hermitian (NH) topological classification approach~\cite{bergholtz2021,kawabata2019}. There, we identify the emergence of inherent NH symmetries~\cite{bernard2002,hui2019}, requiring merely the physical assumptions of thermal equilibrium and Fermi statistics of the constituents. Under these ubiquitous circumstances, pairs of exceptional points (EPs), i.e., NH spectral degeneracies at which the generalized charge susceptibility matrix $\chi ^{\nu \nu'}_c$ becomes non-diagonalizable~\cite{berry2004,heiss2001,heiss2012,heiss2016,shen2018a,kawabata2019b,yang2021,staalhammar2021}, generically occur as a function of a single tuning parameter such as chemical potential (or filling fraction). These EPs are topologically protected by the aforementioned inherent NH symmetry and their splitting in parameter space (see \cref{fig:schematic}(b) for an illustration).

As a consequence, purely real eigenvalues $\lambda_{I}$ of $\chi^{\nu \nu'}_c$ occur in an extended parameter range in between the EPs and generically trigger divergences in the uniform charge response $\chi_{\boldsymbol{q}=0}$~\cite{reitner2020attractive} that are robust against small parameter changes (cf.~\cref{fig:schematic}(a) for an illustration). These divergences in $\chi_{\boldsymbol{q}=0}$ signal the propensity of the correlated system to undergo a thermodynamic phase separation between a compressible metallic and an almost incompressible ``bad metal'' phase, often occurring in the proximity of Mott metal-to-insulator transitions~\cite{kotliar2002compressibility,werner2007mott,eckstein2007,nourafkan2019charge,reitner2020attractive}. This phase separation can be viewed, in many respects, as the electronic counterpart of the liquid gas transition for water molecules.

Non-Hermitian topology and EPs have been widely discussed in systems where the Bloch band structure has been augmented by dissipative terms of various physical origin, ranging from scattering rates of quasi-particles~\cite{yoshida2018,Kimura2019,aquino2019,yoshida2020b,aquino2020, nagai2020,rausch2021,lehmann2021,epfm2023,michen2021,michen2022} to gain and loss in optical systems~\cite{miri2019,weimann2017,takata2018,Zhiyenbayev2019,ashida2020,weidemann2020,menke2017}. There, the experimental visibility of NH signatures is oftentimes limited by the overall blurring introduced by imaginary damping terms in the dissipative time-evolution. By contrast, since the generalized susceptibilities at the heart of our present analysis do not relate to effective complex energy spectra, their NH topology has a direct impact on natural observables, independent of idealistic assumptions on temperature and without the need for complex multi-orbital models, respectively.\\   

\begin{figure}[b]
\centering
\includegraphics[width=0.25\textwidth]{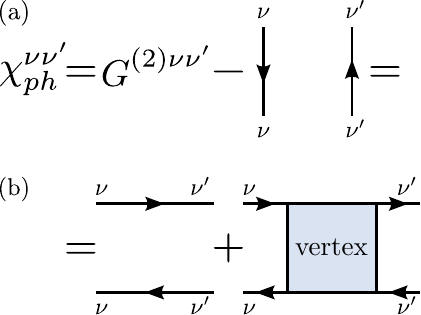}
\hfill
\includegraphics[width=0.2\textwidth]{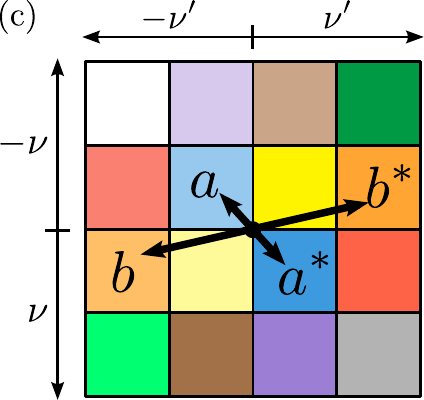}
\caption{\label{fig:matrix} Panels (a) and (b), diagrammatic representation of the generalized susceptibility, as given in \cref{eq:generalChi}, using $\langle \mathcal{T} c^\dagger_{\nu} c^{\pdg}_{\nu'} \rangle = \nu
\fermion
\nu'  = \delta^{\nu\nu'} G(\ii\nu)$
and $\langle \mathcal{T} c^{\pdg}_{\nu } c^{\dagger}_{\nu' } \rangle = \nu
\anfermion
\nu' = -\delta^{\nu\nu'} G(\ii\nu)$, where $G(\ii\nu)$ refers to the one-particle Green's function (here, we dropped the $\alpha_i$ indices for clarity). Panel (c), schematic illustration of the centroHermitian  matrix symmetry.}
\end{figure}

\noindent
{\sl NH symmetries of generalized susceptibilities.} 
The central objects of our analysis are four-point functions describing the propagation of a particle-hole pair (see \cref{fig:matrix}(b)) in the setting of a time-independent Hamiltonian $H$ at thermal equilibrium. 
These are expressed as matrices of two fermionic Matsubara frequencies $\nu$ and $\nu^\prime$, where $\nu^{(\prime)}= (2n^{(\prime)}+1)\pi/\beta$, $n^{(\prime)} \in \mathbb{Z}$, and $\beta=1/T$ is the inverse temperature. 
In the literature, 
such quantities are referred to as \emph{generalized} susceptibilities, since they yield static physical response functions when summed over both fermionic frequencies~\cite{rohringer2018diagrammatic}. 
Specifically, we define them as   
\begin{equation}
        \begin{aligned}
        \chi^{\nu \nu'}_{\ph,\alpha_1 \dots \alpha_4}  = {} &
        \overbrace{\big<\mathcal{T} c^\dagger_{\nu \alpha_1} c^{\pdg}_{\nu \alpha_2} c^\dagger_{\nu' \alpha_3} c^{\pdg}_{\nu' \alpha_4}\big>}^{G^{(2) \nu \nu'}_{\alpha_1 \dots \alpha_4}} \\
        &- \big<\mathcal{T} c^\dagger_{\nu \alpha_1} c^{\pdg}_{\nu \alpha_2}\big>\big<\mathcal{T}c^\dagger_{\nu' \alpha_3} c^{\pdg}_{\nu' \alpha_4}\big>
        \end{aligned}
        \label{eq:generalChi}
\end{equation}
(illustrated in \cref{fig:matrix}(a)), where $\mathcal{T}$ denotes the imaginary time ordering operator, $\big< \dots \big> = 1/Z\Tr(\e^{-\beta H} \dots)$ the thermal expectation value,
$c^{(\dagger)}_{\nu \alpha_i} = \frac{1}{\sqrt{\beta}}\int^\beta_0 \dd \tau \e^{(-)\ii\nu \tau} \e^{H\tau}c^{(\dagger)}_{\alpha_i} \e^{-H\tau}$ the Fourier transform of the (creation) annihilation operators 
\footnote{Here we have shortened the notation of the Fourier transform of the time-ordered operators to make their properties in the following derivations more transparent, in \cref{eq:generalChi} the time ordering is acting first on the time arguments of the operators and then the respective integrals are performed.}, and $G^{(2) \nu \nu'}_{\alpha_1 \dots \alpha_4}$ the two-particle Green's function. 
$\alpha_i$ refers to, in principle, any of the degrees of freedom of the model (momenta, spin, orbital, etc.). 
Some properties of $\chi^{\nu \nu'}_{\ph,\alpha_1 \dots \alpha_4}$ have been already 
analyzed in Refs.~\cite{rohringer2012local,rohringer2014phd, rohringer2018diagrammatic, thunstrom2018analytical,springer2020interplay}. In this work, we are investigating the topological properties of the  
corresponding eigenvalue spectrum.

Taking the complex conjugate of \cref{eq:generalChi} and considering $(c^{\dagger}_\nu)^* = -c^{\pdg}_{-\nu}$, and $(c^{\pdg}_\nu)^* = -c^{\dagger}_{-\nu}$ \emph{inside} of $\big< \dots \big>$,
 one obtains
$(\chi^{\nu \nu'}_{\ph,\alpha_1 \dots \alpha_4})^* = \chi^{-\nu' -\nu}_{\ph,\alpha_4 \dots \alpha_1}$~\footnote{A detailed derivation can be found in the supplemental material~\cite{supplemental}.}.
With simple further manipulations on the indices leaving $\big<\mathcal{T} \dots \big>$ invariant, this leads to
    $(\chi^{\nu \nu'}_{\ph,\alpha_1 \alpha_2 \alpha_3 \alpha_4})^* = \chi^{-\nu -\nu'}_{\ph,\alpha_2 \alpha_1 \alpha_4 \alpha_3}$.
The latter mathematical object has the form of a matrix with coefficients $\chi^{\beta \beta'}_{\ph}$, which, crucially, satisfies the relation
\begin{equation}
\label{eq:compoundChi}
    \chi^{\beta \beta'}_{\ph} = \sum_{\beta_1 \beta_2} \Pi^{\beta \beta_1} (\chi^{\beta_1 \beta_2}_{\ph})^* \Pi^{\beta_2 \beta'},
\end{equation}
where $\Pi^{\beta \beta'}$ is the permutation matrix $\beta:=(\nu,\alpha_1,\alpha_2) \to \beta^{'}:=(-\nu,\alpha_2,\alpha_1)$. This property has important consequences for the eigenvalues $\lambda$ of $\chi_{\ph}$, which belongs to the class of $\kappa$-real matrices $\bm{K}_r= \bm{\Pi} \bm{K}_r^*\bm{\Pi}$~\cite{hill1992k-Hermitian}, where $\bm{\Pi}$ refers to any permutation matrix. These have been shown~\cite{hill1992k-Hermitian} to have a characteristic polynomial with real coefficients and, hence, either real or complex conjugate eigenvalues due to the fundamental theorem of algebra. 
A relevant subclass of $\kappa$-real matrices are \emph{centroHermitian} matrices~\cite{lee1980anna,hill1990Centrohermitian}, which are invariant under a transformation that combines complex conjugation with centro-symmetry, as illustrated in \cref{fig:matrix}(c).

In the following, we consider the \emph{local} generalized charge susceptibility $\chi^{\nu \nu'}_c = \frac{1}{2}\sum_{\sigma \sigma'} \chi^{\nu \nu'}_{\ph,\sigma \sigma \sigma' \sigma'}$ of a one-orbital model that satisfies the following relation:
\begin{equation}
\label{eq:symmetries}
     (\chi^{\nu \nu'}_c)^* = \frac{1}{2}\sum_{\sigma \sigma'} \chi^{-\nu -\nu'}_{\ph,\sigma \sigma \sigma' \sigma'} = \chi^{-\nu -\nu'}_c 
\end{equation}
and is therefore a \emph{centroHermitian} matrix.
In addition, if the Hamiltonian possesses specific symmetries, these can impose even stricter matrix properties. For instance, for particle-hole symmetry (PHS) $\chi^{\nu \nu'}_c$ becomes real and has only real eigenvalues~\cite{thunstrom2018analytical,springer2020interplay}.

\bigskip
\noindent
{\sl Minimal model for exceptional susceptibilities.} To illustrate our general findings, we consider a $2 \times 2$ matrix $ \chi_\ph^{2 \times 2}$ obeying the \emph{centroHermitian} condition:   
\begin{equation}
        \chi_\ph^{2 \times 2} = \left(
        \begin{aligned}
                a + \ii  b && c - \ii d\\
                c + \ii  d && a - \ii b
        \end{aligned}\right)
        = a\cdot \mathbb{I}+\vec{v}\cdot \vec{\sigma}
\label{eq:2times2}
\end{equation}
where $a,b,c,d\in\mathbb{R}$ and $\vec{v}=\vec{v}_{R}+i\vec{v}_{I}=(c,d,0)+i(0,0,b)$ is a complex vector. 
The $a$ parameter can be safely disregarded, as it only amounts to a rigid eigenvalue shift. 
EPs are globally stable for a two- or higher-dimensional parameter space, because for the matrix to become non-diagonalizable, two conditions ($v_{R}^{2}-v_{I}^2=0$, $\vec{v}_{R}\cdot \vec{v}_{I}=0$) have to be simultaneously satisfied. It is immediate to see that the centroHermitian property implies that the second is always fulfilled. It is then sufficient, for the exceptional points to manifest, that $c^2+d^2-b^2=0$, which implies that even in a one-dimensional space, EPs -- if any are present -- will be globally robust against any perturbation representable by a matrix of the form given in \cref{eq:2times2}~\cite{Budich2019}. Crucially, no other perturbation can arise, because the centroHermitian condition does not originate from any further symmetry, but it is an intrinsic consequence of the time-independence of $H$ and a defined quantum (here: Fermi-Dirac) statistics \footnote{See supplemental material for a short discussion on the differences between the susceptibility matrix properties of Fermi-Dirac and Bose-Einstein statistics~\cite{supplemental}.}. The occurrence of these stable EPs generically mark the borders of regions with complex conjugate pairs on the one hand and real and distinct eigenvalues on the other hand.
On the contrary, if we additionally impose PHS, $\chi_\ph^{2 \times 2}$ becomes purely real and symmetric ($b, d = 0$), which implies that the only solution of the two conditions for the EPs will be for $\vec{v}_{R}=\vec{v}_{I}=0$. This is known in literature as a diabolic  point, which is effectively concurrent with a Hermitian degeneracy~\cite{bergholtz2021} and, indeed, generally requires fine-tuning.

\begin{figure}[t]
\includegraphics[width=0.5\textwidth]{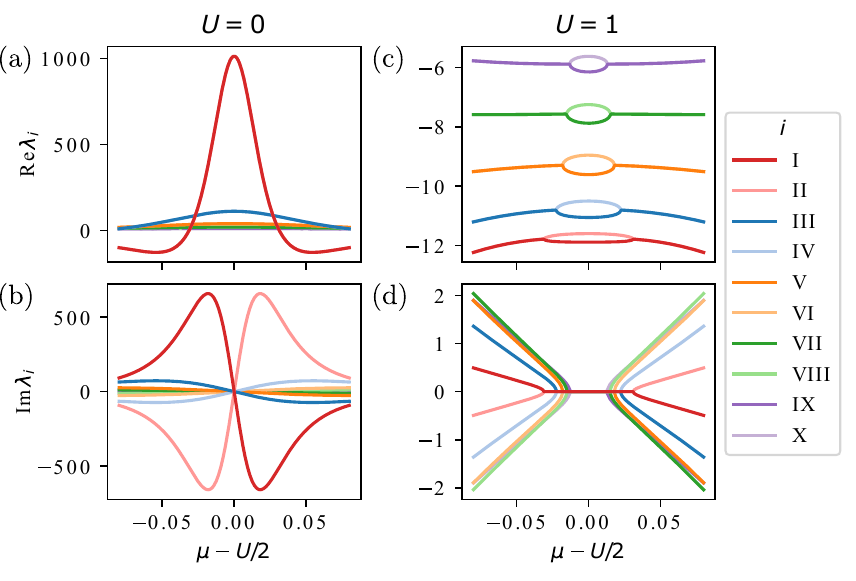}
\caption{\label{fig:alu} Real part (top row) and imaginary part (bottom row) of the eigenvalues $\lambda_i$ of  $\chi^{\nu \nu'}_c$ for the atomic limit (AL) at temperature $T=1/100$, $U=0$ (a, b) and at $U=1$ (c, d) as function of chemical potential away from particle-hole symmetry (PHS) at $\mu=U/2$. The ten eigenvalues which are lowest in $\Re \lambda_i$ at $U=1$ are displayed.}
\end{figure}

\bigskip
\noindent
{\sl Analytical study of the atomic limit.}
As the simplest physical platform to exemplify the spectral properties of $\chi_{c}^{\nu \nu'}$, we now study the exactly solvable atomic limit of the Hubbard model (AL)
\begin{equation}
    H = -\mu (n_\uparrow + n_\downarrow) + U n_\uparrow n_\downarrow,
\end{equation}
where $\mu$ is the chemical potential, $n_\sigma=c^\dagger_\sigma c^\pdg_\sigma$ the occupation of an electron with spin $\sigma$, and $U$ the on-site Coulomb repulsion given in arbitrary units of energy. 
This model fulfills PHS if $\mu = U/2$ and is in general $\operatorname{SU}(2)$-symmetric~\footnote{For the effect of the $\operatorname{SU}(2)$-symmetry on the generalized susceptibilities see Refs.~\cite{bickers2004self, rohringer2012local}.}. In the case of zero interaction $U=0$, the local generalized charge susceptibility reads
\begin{equation}
\label{eq:alu0}
    \chi^{\nu \nu'}_c \stackrel{U=0}{=} -  G(\ii \nu)G(\ii \nu')\delta^{\nu \nu'} = -\frac{\delta^{\nu \nu'}}{(\ii \nu + \mu)^2},
\end{equation}
where $ G(\ii \nu) = \big<\mathcal{T} c^\dagger_{\nu \sigma} c^{\pdg}_{\nu \sigma}\big>$ is the one-particle Green's function. $\chi^{\nu \nu'}_c$ is diagonal, hence the eigenvalues can be immediately read from \cref{eq:alu0}. These become doubly degenerate ($\lambda_\nu=\lambda_{-\nu}=1/\nu^2$) at PHS, i.e. $\mu=0$, while they form complex conjugate pairs ($\lambda_\nu = \lambda_{-\nu}^*$) at finite $\mu$.  
In the left column of \cref{fig:alu} the real (a) and imaginary part (b) of $\lambda_i$ are shown for different $\mu$ at finite temperature $T=1/100$.

At finite interaction $U>0$, $\chi^{\nu \nu'}_c$ becomes a more complicated expression~\cite{thunstrom2018analytical, pairault2000, fus2022bsc, esslARXIV}, given in the supplemental material~\cite{supplemental}. 
The crucial point is the appearance of progressively larger off-diagonal components. The resulting eigenvalues $\lambda_i$ are shown in the right column of  \cref{fig:alu}. Significantly, at PHS, $\lambda_i$ are still purely real but no longer degenerate. 
Importantly, this remains true in a finite region of $\mu$ around $U/2$. 
Far away from PHS, however, the effect of the interaction weakens, and all eigenvalues become complex conjugate pairs. 
To switch between these two regimes, two eigenvalues have to coalesce: this creates a pair of distinct EPs in $\mu$-space, which delimit and protect the real-eigenvalue ``lens''-shaped structure (\cref{fig:alu}(c)).
In the 2$\times$2-picture of \cref{eq:2times2}, we can identify the interaction $U$ as responsible for the presence of the off-diagonal finite elements $c$ and $d$ in the matrix, and the finite $\mu$ for the diagonal element $b$, which are the two ingredients necessary to satisfy the EP conditions. Hence, for the AL any finite value of $U$ will result in exceptional points away from PHS and a finite-size real eigenvalue lens shape.\\

\begin{figure*}[t]
\includegraphics[width=\textwidth]{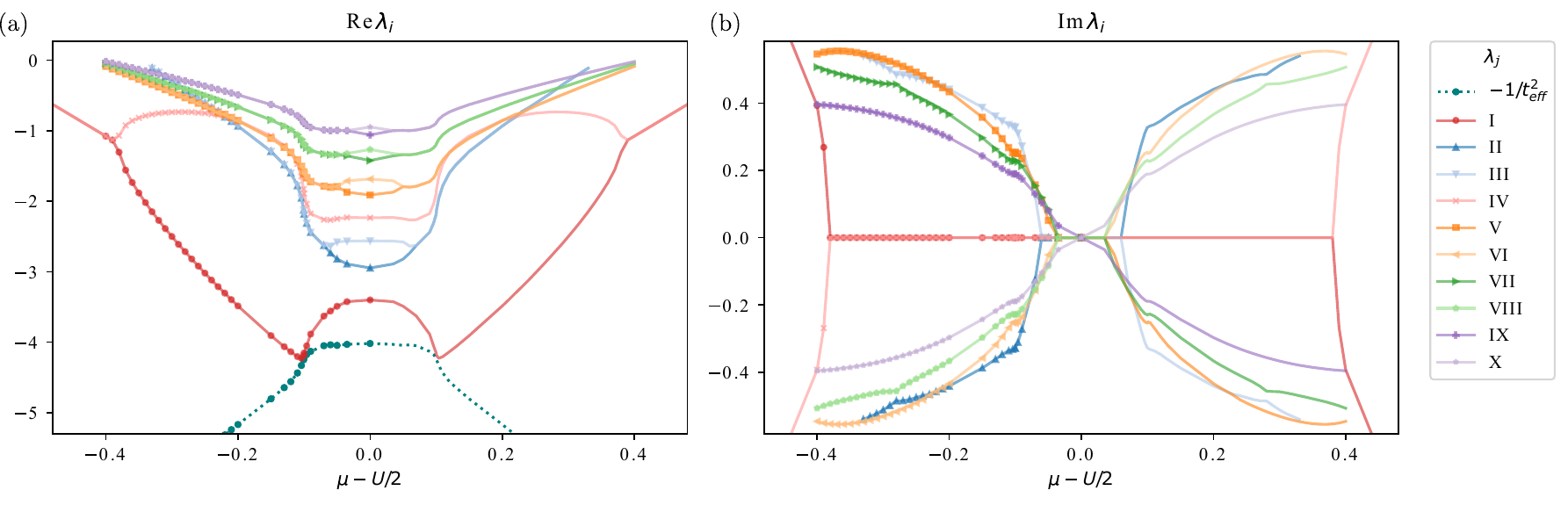}
\caption{\label{fig:dmft} Real part (a) and imaginary part (b) of the eigenvalues $\lambda_i$ of  $\chi^{\nu \nu'}_c$ for the square lattice Hubbard model (with half bandwidth $D = 4t = 1$) solved within DMFT as function of chemical potential away from PHS at
$\mu=U/2$. Interaction strength $U=2.4$ and temperature $T=1/53$ coincide with Ref.~\cite{reitner2020attractive} to show the situation close to the thermodynamic instability. Calculated data are displayed as dots, the positive $(\mu - U/2)$-axis is mapped from the negative one, exploiting the symmetry of the model considered. The ten eigenvalues which are lowest in $\Re \lambda_i$ at $\mu-U/2=0$ are displayed.}
\end{figure*}
\bigskip
\noindent
{\sl Implications on correlation-induced  instabilities.} 
We now turn to a more generic scenario, namely the single-orbital Hubbard Hamiltonian on a lattice:
\begin{equation}
 \begin{aligned}
    \label{eq:hubbard}
    H =& -t \sum_{\langle ij\rangle,\sigma} (c^\dagger_{i \sigma} c^\pdg_{j \sigma} + c^\dagger_{j \sigma} c^\pdg_{i \sigma}) -\mu \sum_{i, \sigma} n_{i \sigma} \\
    &+ U \sum_i  n_{i \uparrow} n_{i \downarrow}.
 \end{aligned}
\end{equation}
with constant hopping $t$ between neighboring sites $i$ and $j$. This model is again $\operatorname{SU}(2)$-symmetric and for $\mu = U/2$ it fulfills PHS. 
Except for one or infinite spatial dimensions, the model has not been solved analytically. In order to get a non-perturbative, albeit approximated many-body solution, we use dynamical mean-field theory (DMFT) (which becomes exact only in the limit of infinite dimensions)~\cite{metzner1989,georges1996dynamical} with a continuous-time quantum Monte Carlo solver from {\it w2dynamics}~\cite{wallenberger2019w2dynamics}.
As shown in Ref.~\cite{reitner2020attractive}, the eigenvalues  $\lambda_i$ and the corresponding eigenvectors $v_i^\nu$  of the local generalized susceptibility $\chi^{\nu \nu'}_c$ ($\sum_{\nu'} \chi^{\nu \nu'}_c  v_i^{\nu'} = \lambda_i v_i^{\nu}$) play an important role for the response functions of the whole lattice: they lead to an enhancement and in some cases to a divergence of the \emph{uniform} (i.e. for zero transfer momentum $\qv=0$) susceptibility. 
In particular, for the Bethe lattice with infinite connectivity (where DMFT is exact) the static uniform charge response, obtained by summing the generalized susceptibility over all Matsubara frequencies $\chi_{\qv=0} = \sum_{\nu \nu'} \chi^{\nu \nu'}_{\qv=0}$, can be re-expressed in terms of $\lambda_i$ and corresponding weights $w_i = (\sum_\nu (v_i^{-1})^{\nu})(\sum_{\nu'} v_i^{\nu'})$. This leads to the following expression
\begin{equation}
\label{eq:DelRe}
    \chi_{\qv=0} = \frac{1}{\beta} \sum_i \left(\frac{1}{\lambda_i} + t^2\right)^{-1} w_i,
\end{equation}
which diverges -- thus inducing a phase instability in the charge sector 
-- when one eigenvalue fulfills the condition $\lambda_i = -1/t^2$. Close to this condition, $\lambda_i$ gives the dominating contribution to the charge response and determines the stability of the physical solution~\cite{kowalski2023thermodynamic}.
Importantly, this is possible only when $\lambda_i$ is \emph{real}~\footnote{Further, the corresponding weight must be $w_i \neq 0$, which is in general fulfilled away from PHS for $\mu \neq U/2$~\cite{reitner2020attractive}}. 

Although \cref{eq:DelRe} is only exact in the case of the Bethe lattice, numerical calculations have shown~\cite{reitner2020attractive} that it holds also for a square lattice if $t$ is replaced by a temperature- and $\mu$-dependent $t_{\mathrm{eff}}(\mu,T)$. 
Here, the central role of EPs becomes apparent: their presence guarantees that the imaginary part of $\lambda_i$ remains zero in the whole extended region of the lens shape. In other words, the possibility of inducing a divergence in $\chi_{\qv=0}$ is not accidental and does not rely on a fine-tuning of $U$, $T$ and $\mu$: the phase instability is in fact  topologically protected. 
For the square lattice, this is illustrated in \cref{fig:dmft}, where we plot the real (a) and imaginary part (b) of the eigenvalues $\lambda_i$ of the local charge susceptibility $\chi^{\nu \nu'}_c$ close to the critical point of the phase separation (cf. sketch in \cref{fig:schematic}). 
Here, the lowest eigenvalue $\lambda_I$ satisfies (up to numerical accuracy) the condition $\lambda_I = -1/t_{\mathrm{eff}}^2$ in the region of the lens shape.
Hence, the phase instability condition is also fulfilled for any further reduction of the temperature $T$ (or for any moderate reduction of the interaction $U$). In particular, at lower $T$, we enter a regime, where a first order phase separation occurs. This regime is, thus, characterized by two locally stable DMFT solutions (i.e., two coexisting values of $\lambda_I$), corresponding to a less correlated metallic and a ``bad metal'' phase (connected by an instable solution, where  $\lambda_I < -1/t_{\mathrm{eff}}^2$~\cite{kowalski2023thermodynamic}). Here, the topological robust arguments related to the condition $\lambda_I = -1/t_{\mathrm{eff}}^2$ remain nonetheless applicable, albeit to the two corresponding metastable solutions~\footnote{A schematic illustration of this regime can be found in the supplemental material~\cite{supplemental}.}. 

Finally, let us notice that a negative eigenvalue is a necessary condition for the instability criterion to be fulfilled~\cite{reitner2020attractive}.  
Remarkably, the role of the negative eigenvalues in the generalized charge susceptibility has been recently related to the local moment formation~\cite{chalupa2021fingerprints,mazitov2022localmomentA,mazitov2022localmomentB,adler2022} and, on a more formal level, to divergences of the irreducible vertex function and the multivaluedness of the Luttinger-Ward functional~\cite{schafer2013divergent,janis2014MITtdivergence,kozik2015nonexistence,stan2015unphysical,rossi2015zero,ribic2016nonlocal,rossi2016shiftedaction,gunnarsson2016parquet,schafer2016nonperturbative,gunnarsson2017breakdown,tarantino2018nonperturbative,vucicevic2018practical,chalupa2018divergences,thunstrom2018analytical,nourafkan2019charge,melnick2020fermi,springer2020interplay,kim2020multivalue,vanloon2020bethesalpeter,reitner2020attractive,chalupa2021fingerprints,kozik2021physical,mazitov2022localmomentA,stepanov2022spindynamics,mazitov2022localmomentB,vanloon2022ipt,adler2022,kim2022misleading,pelz2023highly}.  Therefore, these negative eigenvalues can be regarded as a feature of strong electronic correlations, which cannot be commonly described by ``perturbative'' theories, e.g., the random phase approximation.
However, the considerations behind \cref{eq:DelRe} are not solely restricted to negative eigenvalues. They can be also applied to the opposite case, where a positive eigenvalue reaching a maximum (e.g., $\lambda_i = 1/t_{\mathrm{eff}}^2$) triggers a phase instability, such as the antiferromagnetic transitions of the Hubbard model~\cite{delre2021}. Thus, in these strongly correlated systems, the presence of EPs is found to generally promote phase instabilities in the $\ph$-channel~\footnote{In contrast, for the situation of the generalized  susceptibility in the $\pp$-channel see supplemental material~\cite{supplemental}.} to a stable phenomenon, and thereby enables the instabilities to naturally occur for a finite range in parameter space.
%===============

\bigskip
\noindent
{\sl Conclusion.}
We have found the opening of an EP phase for the associated eigenvalues of the static local susceptibility in the $U/\mu$ phase diagram of models for correlated electron systems. 
The remarkable consequence is that the interaction-induced charge 
instabilities such as the phase separation occurring close to the Mott metal-to-insulator transition in the Hubbard model do not need any fine-tuning but can occur in an entire finite range of parameters.
This unexpected global robustness is a consequence of the peculiar \textit{centroHermitian} form of the susceptibility matrix, which is not dictated by some \textit{ad-hoc} antiunitary symmetry but by the time-independence of $H$ and the intrinsic nature of Fermi-Dirac statistics.

The susceptibility EPs represent a clear-cut and compelling manifestation of non-Hermitian topology, surpassing the conventional realizations based on spectral functions. This phenomenon is indeed ubiquitous even in the simplest correlated fermion models and does not require any assumption on the interaction nor any specific choice of non-Hermitian Hamiltonian terms.
Our results call for future investigations beyond the local correlation effects on the charge sector considered here: e.g., of the spin or particle-particle channel and including non-local correlations in the description. Further, one could also search for higher order exceptional degeneracies in the susceptibility spectrum and explore the respective consequences on the phase instabilities. 
This may open new doors to experimentally detectable hallmarks of non-Hermitian topology.\\

\acknowledgments
\noindent
{\sl Acknowledgments.}
We thank P. Chalupa-Gantner, H. Eßl, P. Oberleitner for insightful discussions, and S. Di Cataldo, P. Kappl, P. Worm for helpful comments. M.R. acknowledges support as a recipient of a DOC fellowship of the Austrian Academy of Sciences and financial support from the Austrian Science Fund (FWF), within the project I 5487. A.T. acknowledges the Austrian Science Fund (FWF) for the project I 5868 (part of the FOR 5249 [QUAST] of the German Science Foundation, DFG).
L.C., J.C.B. and G.S. acknowledge financial support by the Deutsche Forschungsgemeinschaft (DFG, German Research Foundation) under Germany's Excellence Strategy–EXC2147 ``ct.qmat'' (project‐id 390858490) as well as through Project-ID 258499086 – SFB 1170 ``ToCoTronics'' and Project-ID 247310070 - SFB 1143. Calculations have been performed on the Vienna Scientific Cluster (VSC).
\acknowledgments

\bibliography{library} 

\end{document}

\author{M.~Reitner}
\affiliation{Institute of Solid State Physics, TU Wien, 1040 Vienna, Austria}
\author{L.~Crippa}
\affiliation{Institut f\"ur Theoretische Physik und Astrophysik and W\"urzburg-Dresden Cluster of Excellence ct.qmat, Universit\"at W\"urzburg, 97074 W\"urzburg, Germany}
\author{D.~R.~Fus}
\affiliation{Institute of Solid State Physics, TU Wien, 1040 Vienna, Austria}
\author{J.~C.~Budich}
\affiliation{Institute of Theoretical Phyiscs, Technische Universität Dresden and W\"urzburg-Dresden Cluster of Excellence ct.qmat, 01062 Dresden, Germany}
\affiliation{Max Planck Institute for the Physics of Complex Systems, N\"othnitzer Str. 38, 01187 Dresden, Germany}
\author{A.~Toschi}
\affiliation{Institute of Solid State Physics, TU Wien, 1040 Vienna, Austria}
\author{G.~Sangiovanni}
\affiliation{Institut f\"ur Theoretische Physik und Astrophysik and W\"urzburg-Dresden Cluster of Excellence ct.qmat, Universit\"at W\"urzburg, 97074 W\"urzburg, Germany}

\title{Protection of Correlation-Induced Phase Instabilities by Exceptional Susceptibilities\\
{\it -- Supplemental Material --}}

\date{\today}
\maketitle

\noindent
\section{Data availability}
A data set containing all numerical data and plot scripts used to generate the figures of this publication is publicly available on the TU Wien Research Data repository \cite{dataset}.

\noindent
\section{ \label{sm:expl_deriv} Properties of the generalized susceptibility}
The generalized susceptibility in the particle-hole $(\ph)$ channel of the main text explicitly reads
\begin{equation}
 \begin{aligned}
    \chi^{\nu \nu'}_{\ph,\alpha_1 \alpha_2 \alpha_3 \alpha_4} =  \frac{1}{\beta^2 }\int^\beta_0\dd\tau_1\int^\beta_0\dd\tau_2\int^\beta_0\dd\tau_3\int^\beta_0\dd\tau_4 &\\
    \e^{\ii\nu(-\tau_1+\tau_2)}\e^{\ii\nu'(-\tau_3 + \tau_4)}  \Big[&\\
    \frac{1}{Z} \Tr\big(\e^{-\beta H} \mathcal{T}\big\{c^{\dagger}_{\alpha_1}(\tau_1) c^{\pdg}_{\alpha_2}(\tau_2) c^{\dagger}_{\alpha_3}(\tau_3) c^{\pdg}_{\alpha_4} (\tau_4)\big\}\big)&\\
    -\frac{1}{Z} \Tr\big(\e^{-\beta H} \mathcal{T}\big\{c^{\dagger}_{\alpha_1}(\tau_1) c^{\pdg}_{\alpha_2}(\tau_2)\}\big)&\\
    \frac{1}{Z}\Tr\big(\e^{-\beta H} \mathcal{T} \big\{c^{\dagger}_{\alpha_3}(\tau_3) c^{\pdg}_{\alpha_4} (\tau_4)\big\}\big)&\Big],
 \end{aligned}
\end{equation}
where $c^{(\dagger)}_{\alpha_i}(\tau_i)=\e^{H\tau_i}c^{(\dagger)}_{\alpha_i} \e^{-H\tau_i}$. Taking the complex conjugate of this quantity, we get
\begin{equation}
 \begin{aligned}
    \left(\chi^{\nu \nu'}_{\ph,\alpha_1 \alpha_2 \alpha_3 \alpha_4}\right)^* =  \frac{1}{\beta^2 }\int^\beta_0\dd\tau_1\int^\beta_0\dd\tau_2\int^\beta_0\dd\tau_3\int^\beta_0\dd\tau_4 &\\
    \e^{\ii\nu(+\tau_1-\tau_2)}\e^{\ii\nu'(+\tau_3 - \tau_4)} \Big[ &\\
    \frac{1}{Z} \Tr\big(\mathcal{T}\big\{c^{\dagger}_{\alpha_4}(-\tau_4) c^{\pdg}_{\alpha_3}(-\tau_3) c^{\dagger}_{\alpha_2}(-\tau_2) c^{\pdg}_{\alpha_1} (-\tau_1)\big\} \e^{-\beta H} \big)&\\
     -\frac{1}{Z}\Tr\big( \mathcal{T} \big\{c^{\dagger}_{\alpha_4}(-\tau_4) c^{\pdg}_{\alpha_3} (-\tau_3)\big\} \e^{-\beta H}\big)&\\
     \frac{1}{Z} \Tr\big( \mathcal{T}\big\{c^{\dagger}_{\alpha_2}(-\tau_2) c^{\pdg}_{\alpha_1}(-\tau_1)\}\e^{-\beta H}\big)&\Big].
 \end{aligned}
\end{equation}
By changing the integration variables from $-\tau_i \to \tau_i - \beta$
\begin{equation}
 \begin{aligned}
    \left(\chi^{\nu \nu'}_{\ph,\alpha_1 \alpha_2 \alpha_3 \alpha_4}\right)^* =  \frac{1}{\beta^2 }\int^0_\beta\dd\tau_1\int^0_\beta\dd\tau_2\int^0_\beta\dd\tau_3\int^0_\beta\dd\tau_4 &\\
    \e^{\ii\nu(-\tau_1 + \tau_2)}\e^{\ii\nu'(-\tau_3 + \tau_4)}  \Big[&\\
    \frac{1}{Z} \Tr\big(\e^{-\beta H} \mathcal{T}\big\{c^{\dagger}_{\alpha_4}(\tau_4) c^{\pdg}_{\alpha_3}(\tau_3) c^{\dagger}_{\alpha_2}(\tau_2) c^{\pdg}_{\alpha_1} (\tau_1)\big\} \big)&\\
     -\frac{1}{Z}\Tr\big(\e^{-\beta H} \mathcal{T} \big\{c^{\dagger}_{\alpha_4}(\tau_4) c^{\pdg}_{\alpha_3} (\tau_3)\big\}\big)&\\
     \frac{1}{Z} \Tr\big(\e^{-\beta H} \mathcal{T}\big\{c^{\dagger}_{\alpha_2}(\tau_2) c^{\pdg}_{\alpha_1}(\tau_1)\}\big)&\Big],
 \end{aligned}
\end{equation}
where $\e^{\ii \nu^{(\prime)} \beta}=-1$, and interchanging the integration limits
\begin{equation}
 \begin{aligned}
    \left(\chi^{\nu \nu'}_{\ph,\alpha_1 \alpha_2 \alpha_3 \alpha_4}\right)^* =  \frac{1}{\beta^2 }\int^\beta_0\dd\tau_4\int^\beta_0\dd\tau_3\int^\beta_0\dd\tau_2\int^\beta_0\dd\tau_1 &\\
    \e^{-\ii\nu'(-\tau_4 + \tau_3)}\e^{-\ii\nu(-\tau_2 + \tau_1)}  &\\
    \frac{1}{Z} \Tr\big(\e^{-\beta H} \mathcal{T}\big\{c^{\dagger}_{\alpha_4}(\tau_4) c^{\pdg}_{\alpha_3}(\tau_3) c^{\dagger}_{\alpha_2}(\tau_2) c^{\pdg}_{\alpha_1} (\tau_1)\big\} \big) &\\
    -\frac{1}{Z}\Tr\big(\e^{-\beta H} \mathcal{T} \big\{c^{\dagger}_{\alpha_4}(\tau_4) c^{\pdg}_{\alpha_3} (\tau_3)\big\} \big)&\\
    \frac{1}{Z} \Tr\big(\e^{-\beta H} \mathcal{T}\big\{c^{\dagger}_{\alpha_2}(\tau_2) c^{\pdg}_{\alpha_1}(\tau_1)\}\big) &\Big]\\
     =  \chi^{-\nu' -\nu}_{\ph,\alpha_4 \alpha_3 \alpha_2 \alpha_1}&
 \end{aligned}
\end{equation}
we obtain the result of the main text. From the definition of the imaginary time-ordering, it immediately follows
\begin{equation}
     \chi^{-\nu' -\nu}_{\ph,\alpha_4 \alpha_3 \alpha_2 \alpha_1} = \chi^{-\nu -\nu'}_{\ph,\alpha_2 \alpha_1 \alpha_4 \alpha_3}.
\end{equation}
\section{Symmetries}
 Here we recall the properties of $\chi^{\nu \nu'}_{\ph,\alpha_1 \dots \alpha_4}$ under symmetries of the Hamiltonian following Refs.~\cite{rohringer2012local,rohringer2014phd,thunstrom2018analytical,springer2020interplay}. If the Hamiltonian $H$ is invariant under the symmetry transformation $U^\dagger H U^\pdg$, a transformation of all fermionic operators $U^\dagger c^{(\dagger)}_{\alpha_i}(\tau_i) U^\pdg$ will leave $\big< \dots \big>$ invariant~\cite{rohringer2014phd}. E.g., for $\operatorname{SU}(2)$-symmetry it follows from $U^\dagger c^{(\dagger)}_{\sigma_i}(\tau_i) U^\pdg = c^{(\dagger)}_{-\sigma_i}(\tau_i)$ and $U^\dagger H U^\pdg=H$ that 
\begin{equation}
   \chi^{\nu \nu'}_{\ph,\sigma \sigma -\sigma -\sigma} = \chi^{\nu \nu'}_{\ph,-\sigma -\sigma \sigma \sigma} =  \chi^{\nu' \nu}_{\ph,\sigma \sigma -\sigma -\sigma}.
\end{equation}

\section{Local Susceptibilities with Spin Conservation}
In this section, we analyze the matrix properties for the local generalized susceptibilities $\chi^{\nu \nu'}_{\ph,\sigma_1 \dots \sigma_4}$ of a one-orbital model with conservation of spin $\sigma_i$ . This allows us to restrict to the longitudinal ($\chi^{\nu \nu'}_{\ph, \sigma \sigma'} \coloneqq \chi^{\nu \nu'}_{\ph,\sigma \sigma \sigma' \sigma'}$) and transverse ($\chi^{\nu \nu'}_{\ph, \overline{\sigma \sigma'}} \coloneqq \chi^{\nu \nu'}_{\ph,\sigma \sigma' \sigma' \sigma}$) spin components, that satisfy the following two relations:
\begin{subequations}
\begin{eqnarray}
\mathrm{longitudinal}:\; (\chi^{\nu \nu'}_{\ph, \sigma \sigma'})^*  
 & = &\chi^{-\nu' -\nu}_{\ph,\sigma' \sigma} = \chi^{-\nu -\nu'}_{\ph,\sigma \sigma'} \label{subeq:longit} \\
\mathrm{transverse}:\; (\chi^{\nu \nu'}_{\ph, \overline{\sigma \sigma'}})^* & = & \chi^{-\nu' -\nu}_{\ph,\overline{\sigma \sigma'}},
\label{subeq:transv}
\end{eqnarray}
\end{subequations}
and are therefore \emph{centro}- and \emph{perHermitian} matrices, respectively. \emph{PerHermitian} matrices~\cite{hill1990perhermitian} are invariant under a transformation that combines complex conjugation with mirror symmetry with respect to the antidiagonal. They belong to the class of $\kappa$-Hermitian matrices $\bm{K}_h= \bm{\Pi} \bm{K}_r^\dagger\bm{\Pi}$~\cite{hill1992k-Hermitian}, where $\bm{\Pi}$ refers to any permutation matrix, and have  either real or complex conjugate eigenvalues.

\section{Spectral Properties of Centro-/PerHermitian Matrices}
\subsection{CentroHermitian}
\noindent As stated in the main text, a \emph{centroHermitian} matrix $\bm{C}$ has the following property~\cite{lee1980anna,hill1990CentroHermitian} $(C^{\nu \nu'})^*=C^{-\nu -\nu'}$ or equivalently
\begin{equation}
    \bm{J} \bm{C}^* \bm{J} = \bm{C}
\end{equation}
with $J^{\nu\nu'} = \delta^{\nu(-\nu')}$.
Eigenvalues $\lambda$ are either real $\lambda = \lambda^*$ or in complex conjugate (c.c.) pairs $\exists \lambda \rightarrow \exists \lambda^*$. For the eigenvectors $\bm{C}\bm{v} = \lambda \bm{v}$, the following relation  $\exists \bm{v} \rightarrow \exists \bm{Jv}^*$ can be easily shown. For \emph{individual} real eigenvalues, $\bm{v}$ and  $\bm{Jv}^*$ are linear dependent $\rightarrow \bm{v} = \alpha \bm{Jv}^*$ with $\alpha \in \mathbb{C}$.
\begin{align}
    \bm{C}\mathbbm{1}\bm{v} &= \lambda \bm{v}\\
   \bm{J C}^* \bm{JJ v}^* &= \lambda^* \bm{J v}^*\\
  \bm{C J v}^* & = \lambda^* \bm{J v}^*
\end{align}
From this it follows that also the corresponding weights $w$ of the eigenvalues $\lambda$ are either real or appear in  c.c. pairs:
\begin{align}
\begin{split}
    \lambda \in \mathbb{R}:\\
    w_\lambda =& ({\textstyle\sum} \bm{v}) ({\textstyle\sum} \bm{v}^{-1} ) = ({\textstyle\sum}   \alpha \bm{Jv}^*) ({\textstyle\sum}\alpha^{-1} (\bm{v}^{-1})^*\bm{J})\\
    =& ({\textstyle\sum} \bm{v}^*) ({\textstyle\sum} (\bm{v}^{-1})^* ) = w^*_\lambda
\end{split}
\end{align}
\begin{align}
\begin{split}
\lambda, \lambda^* \in \mathbb{C}:\\
    w_{\lambda} =& ({\textstyle\sum} \bm{v}) ({\textstyle\sum} \bm{v}^{-1} ) \\
    w_{\lambda^*} =& ({\textstyle\sum} \bm{Jv}^*) ({\textstyle\sum} (\bm{v}^{-1})^*\bm{J} )\\
    =&  ({\textstyle\sum} \bm{v}^*) ({\textstyle\sum} (\bm{v}^{-1})^* ) = w^*_\lambda
\end{split}
\end{align}
\subsection{PerHermitian}
A \emph{perHermitian} matrix $\bm{P}$ has the following property~\cite{hill1990perhermitian} $(P^{\nu\nu'})^\dagger = P^{-\nu-\nu'}$ or equivalently
\begin{equation}
    \bm{J P}^\dagger \bm{J} = \bm{P}.
\end{equation}
Eigenvalues $\lambda$ are also either real $\lambda = \lambda^*$ or in c.c. pairs $\exists \lambda \rightarrow \exists \lambda^*$. Similar for the centroHermitian case for the eigenvectors $\bm{Pv}=\lambda \bm{v}$ we find the relation 
\begin{align}
    \bm{Pv} &= \lambda \bm{v}\\
    \bm{v}^\dagger \bm{JJ}\bm{P}^\dagger\bm{J} &= \lambda^*\bm{v}^\dagger \bm{J}\\
     \bm{v}^\dagger \bm{J}\bm{P} &= \lambda^*\bm{v}^\dagger \bm{J}.
\end{align}
Hence, for every right eigenvector $\bm{v}$ there exists a left eigenvector $\bm{v}^\dagger\bm{J}$. For \emph{individual} real eigenvalues, this left eigenvector is proportional to the inverse of the right eigenvector $\bm{v}^{-1}=\alpha \bm{v}^\dagger \bm{J}$. From this it again follows that also the corresponding weights $w$ of the eigenvalues $\lambda$ are either real or appear in  c.c. pairs:
\begin{align}
\begin{split}
    \lambda \in \mathbb{R}:\\
    w_\lambda =& ({\textstyle\sum} \bm{v}) ({\textstyle\sum} \bm{v}^{-1} ) = ({\textstyle\sum}\alpha^{-1} \bm{J}(\bm{v}^{-1})^\dagger)({\textstyle\sum}   \alpha \bm{v}^\dagger\bm{J}) \\
    =& ({\textstyle\sum} (\bm{v}^{-1})^* )({\textstyle\sum} \bm{v}^*)  = w^*_\lambda
\end{split}\\
\begin{split}
\lambda, \lambda^* \in \mathbb{C}:\\
    w_{\lambda} =& ({\textstyle\sum} \bm{v}) ({\textstyle\sum} \bm{v}^{-1} ) \\
    w_{\lambda^*} =& ({\textstyle\sum} \bm{J}(\bm{v}^{-1})^\dagger)({\textstyle\sum} \bm{v}^\dagger\bm{J}) \\
    =&  ({\textstyle\sum} (\bm{v}^{-1})^* ) ({\textstyle\sum} \bm{v}^*) = w^*_\lambda
\end{split}
\end{align}
\subsection{Symmetric Centro-/PerHermitian}
\noindent If a centro- or perHermitian matrix $\bm{M}$ is further symmetric $\bm{M}=\bm{M}^T$~\cite{craven1969} we can find an orthonormal eigenbasis $\bm{v}^T\bm{v}=1$ for the subspace where $\bm{M}$ is diagonalizable. For the non-diagonalizable subspaces $\bm{v}$ are  quasi-null $\bm{v}^T\bm{v}=0$. The non-diagonalizable subspace consists of the EPs.

$(\lambda \in \mathbb{R})$:\\
For the orthonormal eigenbasis we can show
\begin{eqnarray}
    \bm{v} &=& \bm{x} + \ii \bm{y}\\
    \bm{v}^T\bm{v} &=& 1 = \bm{x}^T\bm{x} - \bm{y}^T\bm{y} + \ii \underbrace{(\bm{x}^T\bm{y}+\bm{y}^T\bm{x})}_0 \\
     \bm{v}^T\bm{v} &=& 1 = \alpha^2 (\bm{v}^*)^T\bm{JJv}^* = \alpha^2 \bm{v}^T\bm{v},
\end{eqnarray}
thus $\alpha^2=1 \implies \alpha = \pm1$. This means that either the real part $\bm{v}$ is symmetric $\bm{Jx}=\bm{x}$ and the imaginary part is antisymmetric $\bm{Jy}=-\bm{y}$ or vice-versa $\bm{Jx}=-\bm{x}$ and $\bm{Jy}=\bm{y}$. This implies a positive weight  $w>0$ for $\bm{v}=\bm{v}_s$ with a symmetric real part and negative weight  $w<0$ for $\bm{v}=\bm{v}_a$ with an anti-symmetric real part.

\textit{C.c. pairs} $\lambda, \lambda^* \in \mathbb{C}$:\\
For c.c. pairs $\bm{v} \perp \bm{Jv}^*$. Hence, $\bm{v} \neq \pm \bm{Jv}^*$ and $\bm{x}^T\bm{Jx} + \bm{y}^T\bm{Jy} = 0$. This is fulfilled for $\bm{v}_{I}$,$\bm{v}_{II}$ with $\bm{v}_{I} = \frac{1}{\sqrt2}(\bm{v}_1+\bm{v}_2)$ and $v_{II} = \frac{1}{\sqrt2}(\bm{v}_1-\bm{v}_2)$, where $\bm{v}_1 = \pm \bm{Jv}_1^*$ and $\bm{v}_2 = \mp \bm{Jv}_2^*$

\textit{EP}:\\
At an exceptional point (EP) the eigenvector is quasi null $\bm{v}^T\bm{v}=0$ and the eigenvalue $\lambda \in \mathbb{R}$, hence $\bm{x}^T\bm{x}=\bm{y}^T\bm{y}$ and $\bm{v} = \e^{\ii \varphi} \bm{Jv}^*$.

\section{Generalized Charge Susceptibility of the Atomic Limit}
Following Refs.~\cite{thunstrom2018analytical, pairault2000, fus2022bsc} we define the generalized charge susceptibility of the AL in the notation of the main text (with zero bosonic transfer momentum, zero magnetic field, and multiplied with an additional factor of $1/\beta$) as
\begin{widetext}
\begin{equation}
\label{eq:chic_al}
    \begin{aligned}
        \chi^{\nu \nu'}_c =& \phantom{+}\ \frac{(1-\delta^{\nu\nu'}) U^2n(1-n)}{(\ii\nu+\mu)(\ii\nu+\mu-U) (\ii\nu'+\mu)(\ii\nu'+\mu-U)} \\
        &+ \frac{2n-1}{\beta \big(\ii(\nu+\nu')+2\mu-U\big)}\left(\frac{1}{\ii\nu+\mu-U}+\frac{1}{\ii\nu'+\mu-U}\right)^2 \\
        &- \delta^{\nu\nu'}\frac{\e^{\beta \mu}}{Z}\left(\frac{1}{\ii\nu+\mu}-\frac{1}{\ii\nu'+\mu-U}\right)^2\\
        &+ \frac{\e^{\beta(2\mu-U)}-\e^{\beta2\mu}}{Z^2}\, \frac{U^2 }{(\ii\nu+\mu)(\ii\nu+\mu-U) (\ii\nu'+\mu)(\ii\nu'+\mu-U)}\\
        &+ \frac{1-n}{\beta(\ii\nu+\mu)^2}\left(\frac{1}{\ii\nu'+\mu}-\frac{1}{\ii\nu'+\mu-U}\right) + \frac{1-n}{\beta(\ii\nu'+\mu)^2}\left(\frac{1}{\ii\nu+\mu}-\frac{1}{\ii\nu+\mu-U}\right)\\
        &+ \frac{1}{\beta(\ii\nu+\mu-U)^2}\left(\frac{1-n}{\ii\nu'+\mu} -\frac{n}{\ii\nu'+\mu-U}\right) + \frac{1}{\beta(\ii\nu'+\mu-U)^2}\left(\frac{1-n}{\ii\nu+\mu}-\frac{n}{\ii\nu+\mu-U}\right)\\
        &+\frac{2(1-n)}{\beta(\ii\nu'+\mu)(\ii\nu'+\mu-U)}\left(\frac{1}{\ii\nu+\mu-U}-\frac{1}{\ii\nu+\mu}\right)\\
        &-\delta^{\nu \nu'} \left(\frac{1-n}{\ii\nu+\mu} +\frac{n}{\ii\nu+\mu-U}\right)^2,
    \end{aligned}
\end{equation}
\end{widetext}
where $Z=1+2\e^{\beta \mu} + \e^{\beta(2\mu-U)}$ and $n=(\e^{\beta\mu}+\e^{\beta(2\mu-U)})/Z$. Note that the term in the second line of \cref{eq:chic_al} becomes proportional to $\delta^{\nu \nu'}$ for PHS at $\mu=U/2$.

\section{$\pp$-Channel}
For the \emph{local} generalized susceptibility $\chi^{\nu \nu'}_{\pp,\sigma_1 \dots \sigma_4}$ of a one-orbital model with conservation of spin $\sigma_i$ the longitudinal particle-particle ($\pp$) channel has quite different spectral properties compared to the $\ph$-channel. $\chi^{\nu \nu'}_{\pp,\sigma_1 \dots \sigma_4}$ is defined as
\begin{equation}
 \begin{aligned}
        \chi^{\nu \nu'}_{\pp,\sigma_1 \dots \sigma_4} = {}&\ 
        \big<\mathcal{T} c^\dagger_{\nu \sigma_1} c^{\pdg}_{-\nu' \sigma_2} c^\dagger_{-\nu \sigma_3}  c^{\pdg}_{\nu' \sigma_4}\big>\\
        &- \big<\mathcal{T} c^\dagger_{\nu \sigma_1} c^{\pdg}_{-\nu' \sigma_2}\big>\big<\mathcal{T}c^\dagger_{-\nu \sigma_3} c^{\pdg}_{\nu' \sigma_4}\big>.
 \end{aligned}
\end{equation}
Again taking its complex conjugate and considering spin conservation we obtain two relations:
\begin{subequations}
\label{eq:pp-symmetries}
\begin{eqnarray}
    \mathrm{longitudinal}: (\chi^{\nu \nu'}_{\pp,\sigma \sigma'})^* &=& \chi^{-\nu' -\nu}_{\pp,\sigma' \sigma} = \chi^{\nu' \nu}_{\pp,\sigma  \sigma'}  \label{subeq:pp-longit} \\
    \mathrm{transverse}: (\chi^{\nu \nu'}_{\pp,\overline{\sigma \sigma'}})^* &=& \chi^{-\nu' -\nu}_{\pp,\overline{\sigma \sigma'}} \label{subeq:pp-transv}.
\end{eqnarray}
\end{subequations}
The transversal spin components $\chi^{\nu \nu'}_{\pp,\overline{\sigma \sigma'}}$ in the $\pp$-channel form a \emph{perHermitian} matrix, however \eqref{subeq:pp-longit} shows that the longitudinal components $\chi^{\nu \nu'}_{\pp,\sigma \sigma'}$ now make up a \emph{Hermitian} matrix and EPs are not required to ensure real eigenvalues $\lambda$ for an instability condition.\\

\section{Instability condition in the region of phase separation}
\begin{figure}[h!]
\includegraphics[width=0.5\textwidth]{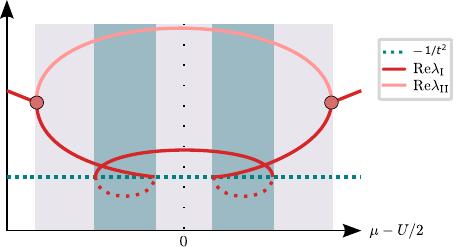}
\caption{\label{fig:coexistence} Schematic illustration of the fulfillment of the instability condition in the region of the phase separation (turquoise background). The real parts of the lowest eigenvalue $\lambda_I$ and the associated eigenvalue $\lambda_{II}$, for which  $\lambda_I$ and $\lambda_{II}$ coalesce at the exceptional points (red dots), are displayed. $\lambda_I$ is shown as dotted line below the limit $-1/t^2$ to indicate the instability of this solution. In the region of the ``lens shape'' (gray background) both eigenvalues have a zero imaginary part.}
\end{figure}

In \cref{fig:coexistence}, we schematically illustrate the fulfillment of the instability condition in the region of the phase separation (turquoise background). Here, two locally stable DMFT solutions (i.e., two coexisting values of $\lambda_I$ ), corresponding to a less correlated metallic and a ``bad metal'' phase are connected by an unstable solution (red dotted line), where $\lambda_I<-1/t^2$~\cite{kowalski2023thermodynamic}.
The instability condition $\lambda_I=-1/t^2$ is fulfilled in the region of the ``lens shape'' (gray background) between the exceptional points, where $\lambda_I$ is real. For the two locally stable DMFT solutions of $\lambda_I$, the one which meets the instability criterion corresponds to the metastable solution (the other is thermodynamically stable).

\section{Generalized susceptibility matrix properties for bosons}
By taking bosonic $b^{(\dagger)}$ instead of fermionic $c^{(\dagger)}$ operators for the generalized susceptibility we get
\begin{equation}
        \begin{aligned}
        \chi^{\omega \omega'}_{\ph,\alpha_1 \dots \alpha_4}  = {} &
       \big<\mathcal{T} b^\dagger_{\omega \alpha_1} b^{\pdg}_{\omega \alpha_2} b^\dagger_{\omega' \alpha_3} b^{\pdg}_{\omega' \alpha_4}\big> \\
        &- \big<\mathcal{T} b^\dagger_{\omega \alpha_1} b^{\pdg}_{\omega \alpha_2}\big>\big<\mathcal{T}b^\dagger_{\omega' \alpha_3} b^{\pdg}_{\omega' \alpha_4}\big>
        \end{aligned}
        \label{eq:bosonChi}
\end{equation}
where
$b^{(\dagger)}_{\omega \alpha_i} = \frac{1}{\sqrt{\beta}}\int^\beta_0 \dd \tau \e^{(-)\ii\omega \tau} \e^{H\tau}b^{(\dagger)}_{\alpha_i} \e^{-H\tau}$ are now the Fourier transforms of bosonic (creation) annihilation operators. The corresponding generalized susceptibility is then describing the propagation (and scattering) of two Bosons instead of two Fermions. The crucial difference of \cref{eq:bosonChi} to Eq.~(1) in the main text are the bosonic (even) Matsubara frequencies  $\ii \omega^{(\prime)} = 2 n^{(\prime)} \pi/\beta$, $n^{(\prime)} \in \mathbb{Z}$. For $ \chi^{\omega \omega'}_{\ph,\alpha_1 \dots \alpha_4}$ we can also follow the steps in section ``Properties of the Generalized Susceptibility'' of this Supplemental Material (now $\e^{\ii\omega^{(\prime)}\beta}=1$ ) and arrive at the same relation we have found for Fermions:
\begin{equation}
      \left( \chi^{\omega \omega'}_{\ph,\alpha_1 \alpha_2 \alpha_3 \alpha_4}\right)^*=  \chi^{-\omega -\omega'}_{\ph,\alpha_2 \alpha_1 \alpha_4 \alpha_3}.
\end{equation}
Hence, for Bosons the corresponding $\chi^{\beta \beta'}_{\ph}$ matrix also satisfies the relation
\begin{equation}
\label{eq:compoundChi}
    \chi^{\beta \beta'}_{\ph} = \sum_{\beta_1 \beta_2} \Pi^{\beta \beta_1} (\chi^{\beta_1 \beta_2}_{\ph})^* \Pi^{\beta_2 \beta'},
\end{equation}
where $\Pi^{\beta \beta'}$ is the permutation matrix of $\beta:=(\omega,\alpha_1,\alpha_2) \to \beta^{'}:=(-\omega,\alpha_2,\alpha_1)$, and $\chi^{\beta \beta'}_{\ph}$ has the same properties for its eigenvalues $\lambda$: they are either real or complex conjugate pairs. 

However, although the general matrix property remains the same, the choice of fermionic or bososnic Matsubara frequencies does have an important implication on the eigenspectrum of the matrix. The crucial difference is the presence of the zero frequency $\ii\omega=0$ among the bosonic Matsubara frequencies which is absent for fermionic ones. In addition to the expected complex conjugate eigenvalue pairs and EPs of the fermionic system, the bosonic system has, thus, an extra eigenvalue, which remains always real, even without further symmetries.

Let us then assume for a moment that a similar mechanism, as the one we investigated for the fermionic system, might be also responsible for a phase instability of the bosonic system. Then, since such an instability requires a real eigenvalue to reach a certain threshold, EPs might or might not play a crucial role, however they are no longer a necessary condition for the fulfillment of the instability, due to the extra real eigenvalue.

\begin{figure}[t]
\includegraphics[width=.5\textwidth]{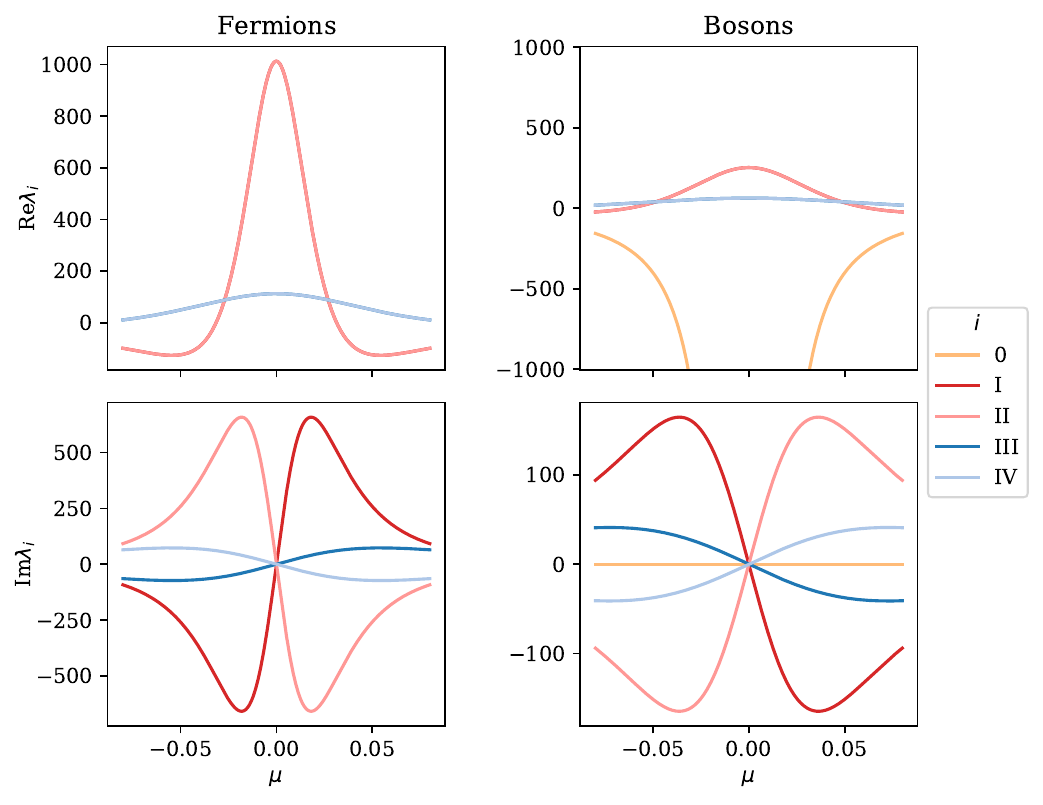}
\caption{\label{fig:bosonic} Real part (top row) and imaginary part (bottom row) of the eigenvalues $\lambda_i$ of the generalized charge susceptibility   for the non-interacting ($U=0$) atomic limit at temperature $T=1/100$ for Fermions $\chi^{\nu \nu'}_c$ (left column) and Bosons $\chi^{\omega \omega'}_c$ (right column) as function of chemical potential. The four resp.~five eigenvalues stemming from the inner Matsubara frequencies are displayed. The positive (non physical) values of the chemical potential $\mu$ for the bosonic system are included for completeness. }
\end{figure}

Consider the simple example of the non-interacting susceptibility of the atomic limit for Fermions
\begin{eqnarray}
    \chi^{\nu \nu'}_F \stackrel{U=0}{=} -  G(\ii \nu)G(\ii \nu')\delta^{\nu \nu'} = -\frac{\delta^{\nu \nu'}}{(\ii \nu + \mu)^2},
\end{eqnarray}
 where $\frac{1}{\beta}\sum_{\nu \nu'}\chi^{\nu \nu'}_F = \partial_\mu n_F(-\mu)$, and for Bosons
\begin{eqnarray}
    \chi^{\omega \omega'}_B \stackrel{U=0}{=} -  G(\ii \omega)G(\ii \omega')\delta^{\omega \omega'} = -\frac{\delta^{\omega \omega'}}{(\ii \omega + \mu)^2},
\end{eqnarray}
 where $\frac{1}{\beta}\sum_{\omega \omega'}\chi^{\omega \omega'}_B = \partial_\mu n_B(-\mu)$ ($n_F(\epsilon)=1/(\e^{\beta\epsilon}+1)$ and  $n_B(\epsilon)=1/(\e^{\beta\epsilon}-1)$). The resulting eigenvalue spectrum $\lambda_i$ is displayed in Fig.~\ref{fig:bosonic}.  By comparing the spectra of the bosonic (right column) with the fermionc system (left column) an additional eigenvalue $\lambda_0=-1/\mu^2$ can be found, which then remains real also out of particle-hole symmetry independent of the value of the chemical potential $\mu$. Interestingly in this simple example, $\lambda_0$ is the eigenvalue responsible for the (negative) divergence of $\partial_\mu n_B$ at $\mu=0$, generically associated to the possible onset of Bose-Einstein condensations.

\bibliography{library}